# A Secure and Authenticated Key Management Protocol (SA-KMP) for Vehicular Networks

Heng Chuan Tan, Maode Ma, *Senior Member*, IEEE, Houda Labiod, *Member*, IEEE, Aymen Boudguiga, Jun Zhang, *Member*, IEEE, Peter Han Joo Chong, *Member*, IEEE

*Abstract*— **Public Key Infrastructure (PKI) is the most widely used security mechanism for securing communications over the network. However, there are known performance issues, making it unsuitable for use in the Vehicular Networks (VNs). In this paper, we propose a Secure and Authenticated Key Management Protocol (SA-KMP) to overcome the shortcomings of the PKI. The SA-KMP scheme distributes repository containing the bindings of the entity's identity and its corresponding public key to each vehicle and Road Side Unit (RSU). By doing so, certificate exchanges and Certificate Revocation Lists (CRLs) are eliminated. Furthermore, SA-KMP scheme uses symmetric keys derived based on a 3D matrix based key agreement scheme to reduce the high computational costs of using asymmetric cryptography. We demonstrate the efficiency of the SA-KMP through performance evaluations in terms of transmission and storage overhead, network latency and key generation time. Analytical results show that the SA-KMP is more scalable and outperforms the certificate based PKI. Simulation results indicate that the key generation time of the SA-KMP scheme is less than that of the existing Elliptic Curve Diffie-Hellman (ECDH) and Diffie-Hellman (DH) protocols. In addition, we use Proverif to prove that the SA-KMP scheme is secure against an active attacker under the Dolev and Yao model and further show that the SA-KMP scheme is secure against Denial of Service (DoS), collusion attacks and a wide range of other malicious attacks.**

## I. INTRODUCTION

In recent years, there has been a change of research emphasis from Vehicular Ad-hoc Networks (VANETs) to intelligence vehicles which lead to the development of Cooperative Intelligence Transport System (C-ITS). In C-ITS, all the elements of the transport chain ranging from the public transport down to the road users are connected to form a Vehicular Network (VN). It aims to improve the public road safety and optimize the traffic management [1]-[3]. In general, a VN comprises of two network entities: road users and Road Side Units (RSUs). The road users are vehicles, pedestrians, motorized cyclists, etc. traveling at different speeds while the RSUs are statically deployed to form the backbone of the network. In such a configuration, different modes of communications can be realized, namely the Vehicle to Infrastructure (V2I), Infrastructure to Vehicle (I2V), Vehicle to Vehicle (V2V) and Vehicle to Pedestrians (V2P). To enable these devices to communicate, each entity is equipped with an On-Board Unit (OBU) that employs the Dedicated Short Range Communications (DSRC) technology [4]. Besides the safety and traffic optimization applications, VNs have also evolved to become a service provider to support value-added services such as the infotainment applications and Internet access provisioning.

To support these diverse applications, an efficient key management system is needed to ensure secure and reliable exchanges of information. A traditional approach is to use a certificate-based Public Key Infrastructure (PKI) as recommended in the IEEE 1609.2 [5] and the ETSI TS 102 940 standards [6]. The PKI uses the public key cryptography with digital certificates to provide confidentiality, authenticity, integrity, and non-repudiation. However, several issues impede the deployment of PKI [7] in the VNs. First of all, a vehicle needs to distribute its public key certificate signed by the Certificate Authority (CA) to other users. This results in the wastage of communication bandwidth. After receiving the certificate, each entity has to (1) verify the expiry date of the certificate, (2) check the validity of the certificate against a Certification Revocation List (CRL) and (3) verify the CA's digital signature on the signed certificate. These requirements introduce much latency which is undesirable for most VNs applications. It is also necessary to distribute a huge CRL, which does not scale well with increasing network size. Moreover, if the CRLs are not disseminated on time, a recipient will be at risks of accepting an expired or previously revoked certificate [7].

Motivated by the shortcomings of the PKI, we propose an efficient Secure and Authenticated Key Management Protocol (SA-KMP) to overcome the complex verification process inherent in all the certificate-based PKI schemes. Our work combines two prior works, namely the Public Key Regime (PKR) [18] and the 3D matrix key distribution scheme [10], [11]. The PKR eliminates the exchange of digital certificates by delegating the distribution of public keys to the RSUs while the

H C Tan, and M. Ma are with the Nanyang Technological University, Nanyang Avenue, Singapore (e-mail: htan005@ntu.edu.sg, emdma@ntu.edu.sg). H C Tan is also with INFRES, Telecom ParisTech, 46, rue Barrault, 75013, Paris cedex 13, France.

H. Labiod and J. Zhang are with INFRES, Telecom ParisTech, 46, rue Barrault, 75013, Paris cedex 13, France. (e-mail: houda.labiod@telecom-paristech.fr, jun.zhang@telecom-paristech.fr).

A. Boudguiga is with IRT SystemX,8 avenue de la Vauve - CS, 90070, 91127 Palaiseau CEDEX, France, (e-mail: aymen.boudguiga@irt-systemx.fr)

P.H.J. Chong is with Auckland University of Technology, Auckland, New Zealand, (e-mail: peter.chong@aut.ac.nz).

3D matrix key scheme generates common keys for use in symmetric encryption to reduce the dependence on expensive asymmetric operations. The major contributions of our work are summarized below:

1) Develop a messaging protocol for negotiating pairwise keys in a V2I/I2V and V2V communication.
2) Develop a novel key distribution scheme based on the 3D matrix key scheme to generate the keys dynamically instead of preloading the keys.
3) Incorporate an authentication mechanism into the PKR scheme and key distribution scheme to mitigate Denial of Service (DoS) attacks.

The rest of the paper is organized as follows: Section II reviews the related work. Section III describes the system model and highlights our design considerations. Section IV discusses the mechanisms and the framework of the SA-KMP. Section V discusses the security properties of our scheme. Section VI analyzes the performance of our scheme. Section VII concludes the paper.

## II. RELATED WORKS

There are many methods to reduce the high costs of deploying a PKI which we classify into three types. The first approach is to secure the messages using symmetric cryptography rather than asymmetric cryptography. By the symmetric cryptography, the communicating parties have to know the secret key before encryption. Thus, several key distribution schemes have been proposed in the literature to distribute the keys securely. The second method is to avoid the use of digital certificates entirely so that expensive operations related to certificate management are eliminated. The third method is by reducing the CRL checking delay.

### A. Key Distribution Schemes

In [12], J. Almeida et al. have proposed a probabilistic key sharing scheme where each vehicle is given a key ring containing $k$ keys drawn randomly from a large pool of $P$ keys. For any two vehicles to determine a common key, each of them has to broadcast the list of the key identifiers on their key ring using a challenge-response protocol. In [13], D.A. Don et al. have introduced the multivariate polynomial scheme by which two vehicles establish a key by substituting the unique ID into its own univariate polynomial share pre-loaded by the authority. Since the univariate polynomials are extracted from a symmetric bivariate polynomial of degree $t$, it follows that two communicating vehicles can derive the same common keys. In [14] and [15], Y. Zhang et al. have proposed to establish a pairwise key based on the symmetry properties of the matrix operations. The authority first creates a public matrix $G$ and a secret symmetric matrix $D$ and then generates another public matrix A which is the transpose of the matrix $G$ and $D$. The two communicating parties establish a common key by multiplying the two public matrices together. In [16], HT. Wu et al have proposed to use the Diffie-Hellman (DH) protocol to establish a secret key. The secret key is then passed as an input into a hash chain to generate as many session keys as the number of RSUs in the region. In [17], a distributed group key agreement scheme is proposed based on the Elliptic Curve Diffie-Hellman (ECDH) protocol which is adaptable to membership changes in a group. In [10] and [11], M.A. Hamid et al have proposed a 3D matrix key distribution scheme by which both parties solve a system of plane equations to identify the common intersection points they share. The key preloaded at each intersection point is then used to secure the communications. By all of the abovementioned schemes, the public key cryptography is required to protect the exchange of key information for deriving the common keys which means that an adversary could launch DoS attacks by sending a lot of bogus messages for verifications. Additionally, the schemes in [10]-[12] are susceptible to the node capture attacks because the adversary can compromise a node physically to extract the preloaded keys.

### B. Certificate-less PKI

In [18], P.Y. Shen et al. have proposed the Public Key Regime (PKR) scheme to reduce the cost of managing certificates. Each RSU is given a read-only copy of a public key directory by the Trusted Authority (TA). The public key directory contains the vehicle identifiers and their corresponding public keys that have previously been certified by a TA. By distributing the public key directory to the RSUs, it eliminates the exchange and verification of the certificates. However, the PKR is not secured against the DoS and collusion attacks since a malicious vehicle can send many public key request messages to overwhelm the RSU.

### C. Simple CRL Check

In [19], A.H. Salem et al. have introduced the RSU managers into the PKI hierarchy to keep track of the RSUs that assist the requesting vehicles. This scheme can improve the revocation process because the revocation message is only sent to the RSU that has the highest probability of contacting the moving vehicle. In [20], an Expedite Message Authentication Protocol (EMAP) has been proposed based on the keyed-Hash Message Authentication Code (HMAC) with the probabilistic key sharing scheme. The HMAC is verified first before the digital certificate is checked thereby improving the CRL checking process. On the other hand, the probabilistic key sharing mechanism helps to update the key used for the HMAC. However, by both schemes, a CRL is still required to distribute certification revocation information which increases the communication overhead.

In this paper, SA-KMP is developed based on integrating the PKR scheme [18] and the 3D matrix key scheme [10], [11] together. Using these two concepts as a basis, we propose further extensions to provide resilience against DoS attacks and node capture attacks. Table I highlights the novelty of the SA-KMP scheme by comparing it with the existing works.

## III. SYSTEM MODEL

In this section, we describe the network model, the communication model, the security model, our design considerations and the assumptions used.

TABLE I
COMPARISON VERSUS RELATED WORKS

| Scheme | Mechanism | (A) | (B) | (C) |
|---|---|---|---|---|
| Probability based[12] | Preload | x | x | x |
| Polynomial based[13] | Dynamic key generation | x | √ | x |
| Matrix based[14], [15] | Dynamic key generation | - | √ | x |
| DH based[16], [17]] | Dynamic key generation | x | √ | x |
| 3D matrix based[10],[11] | Preload | - | x | x |
| PKR [18] | Certificate-less | √ | - | x |
| [19] | Reduce CRL check | x | - | x |
| [20] | Reduce CRL check | x | - | x |
| **Proposed SA-KMP** | **Certificate-less + Dynamic key generation** | √ | √ | √ |

(A) Certificate-less, (B) Resilient against Node Capture, (C) Resilient against DoS attack

*A. Network Model*

We consider a VN deployed in an urban area consisting of three entities:

**(1) Regional Transport Authority (RTA) -** The RTA is in charge of the management of the vehicles and RSUs in a geographical area. It is responsible for the registration of vehicles and RSUs, issuance of private/public key pairs, creation of the public repositories including updating and dissemination of the repositories. It is fully trusted by all the vehicles and RSUs in the deployment area and is equipped with advanced security mechanisms to prevent any information leakage. Thus, it has the highest level of security and cannot be compromised.

**(2) Road Side Units (RSUs) -** The RSUs are statically and strategically deployed to bridge the communication among the vehicles or RSUs in the network. They are installed at traffic lights, lamp posts or road signs, etc. They are connected to the RTA via a secure network and are equipped with multiple interfaces for interoperability with different access technologies. We assume that the RSUs are supported by the location-based services that enable them to know their location and the locations of other entities within their communication range. They take part in the pairwise key agreement for V2I/I2V and V2V communications. It is assumed that the RSUs have no resource constraints regarding storage and computing power and that the majority of the RSUs is trusted.

**(3) Vehicles -** The vehicles are either static or mobile, traveling on the network at various speeds. They are equipped with OBUs to support communications with other vehicles and RSUs. The OBU of a vehicle is also installed with the Global Positioning Service (GPS) to provide useful information about the car's position and the position of other entities on the road. We assume that some vehicles are not trusted.

*B. Communication Model*

We consider two types of communication in the VNs, namely the V2I and V2V. The V2I mode refers to the communication between the RSU and the vehicle and vice versa. An example of V2I is the periodic broadcast of beacon messages containing the vehicle's speed or position information to the RSU whereas I2V refers to the sending of safety related messages from the RSU to the vehicles. V2V, on the other hand, refers to the ad-hoc communication among the vehicles on the road. They exchange information such as warning messages to alert other drivers on the accidents or collisions ahead. To secure these two types of communication, we define a pairwise key to be shared between two entities in a V2I or V2V communication.

*C. Attack and Security Model*

Attacks can be classified into two types: passive and active attacks. The passive attackers monitor the communication channel to gather sensitive information without modifying the information. On the other hand, the active attackers as described by the Dolev and Yao attacker model [21], may intercept messages, alter them or replay the original messages to gain access to confidential information. The attackers can also act irrationally if the benefits outweigh the risk of being detected. Interested readers may refer to [2], [8] and [22] for a comprehensive coverage of the threats and the attacks. In this paper, we consider the presence of active attackers such as the DoS and collusion attacks, which can be launched by the attackers when the vehicles request for public keys from the RSU and vice versa. The DoS attacks can also occur during the key agreement process. In terms of security capability, we assume that the vehicles and RSUs are fitted with a Hardware Security Module (HSM) responsible for storing, and physically protecting the cryptographic information similar to the implementation in [22]. Confidential information stored in the HSM includes node ID, node's private key/public key, RTA's public key, session keys and some cryptomaterial used for establishing the pairwise keys. In addition, the HSM is responsible for performing cryptographic key operations during key establishment.

*D. Design Considerations*

The SA-KMP scheme is developed based on the following considerations. The SA-KMP has to be **efficient** due to the strict delay requirements of VN applications. Distribution and verification of public keys cannot be too complex to reduce the high overhead and delay. At the same time, it should not compromise on the level of security. It has to be **lightweight** due to the limited computing capability of the vehicles as compared to the RSU. As the number vehicles increases, the number of certificates or keys will also increase. Therefore, it is important that our scheme is **highly adaptable and scalable** to the growth of the network. Furthermore, all the messages in the V2I or V2V communications must be protected against any data modifications to ensure information **integrity**. If messages are modified intentionally, it could lead to adverse consequences such as loss of lives. Recipients of a message must also be able to verify the **authenticity** of the message, that is, the message is sent by an authorized entity with the proper signing key. The SA-KMP must also be able to guarantee **non-repudiation** property where the sender or receiver of a message cannot deny having sent or received the message. Lastly, the information in a message must be kept secret and protected from unauthorized access at all times to ensure data **confidentiality**.

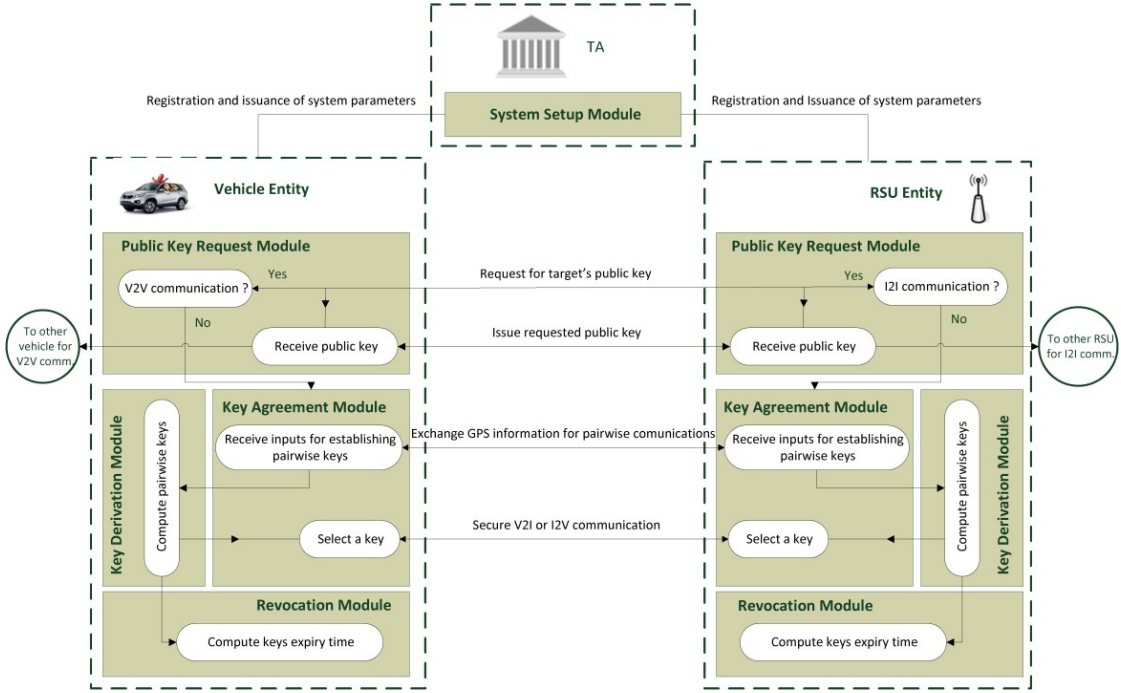

Fig. 1. SA-KMP Framework

## IV. THE SA-KMP FRAMEWORK

Figure 1 describes the various modules of the SA-KMP scheme and their interactions with external entities such as the RSU and the vehicles. The five modules are the System Setup, Public Key Request, Key Agreement, Key Derivation and Revocation. Table II defines the notations used in this paper.

### A. System Setup Module

This module is executed by the RTA and consists of 3 steps.

**Step 1 (Initialization):** The deployment area is divided into a 3D space ($m \times m \times m$) where $m$ represents the size of the area. Each entity occupies a unique location denoted by $(i, j, k)$ in the 3D space. The RTA determines the domain parameters $(p, a, b, G, n, h)$ for the elliptic curve $E$ defined over the finite field, $F_p$ where $p$ is a large prime number and $a$, $b$ are the parameters that define the elliptic curve $E$ of the form $y^2 \pmod{p} = x^3 + ax + b \pmod{p}$. $G$ is the generator denoted by a point $(G_x, G_y)$ chosen from the elliptic curve, $n$ is the order of the generator, $h$ is the cofactor and $h: \{0,1\}^* \to F_p$ is a secure hash function. All the system parameters $(p, a, b, G, n, h)$ are made public to all the entities in the network. The RTA then generates a set of $N$ plane equations having this form given in (1).

$$\begin{aligned} z - k_i + c_1(y - j_i) + c_2(x - i_i) &= 0 \bmod(m) \\ z - k_i + c_3(y - j_i) + c_4(x - i_i) &= 0 \bmod(m) \\ z - k_i + c_5(y - j_i) + c_6(x - i_i) &= 0 \bmod(m) \\ \forall\, c_1, c_2, \cdots, c_N \end{aligned} \quad (1)$$

where $(i_i, j_i, k_i)$ denotes the GPS location of the entity and the subscript $i$ represents the ID of the entity. The parameter, $m$ denotes the size of the 3D space and $c_1, c_2, \cdots, c_N$ are constants of the plane equations where $c_1 \neq c_2 \neq \cdots \neq c_6 \neq c_N$. The constants are selected such that the rank of the coefficient matrix is of full rank. It guarantees that there is always a unique solution to a system of three plane equations. Next, the RTA generates another three hash functions ($H_1$, $H_2$, $H_3$) which are used for deriving the common keys and creates two directories, namely the RSU-Public File Directory (RSU-PFD) and the Vehicle Public Directory (VPFD) to store the entities' public keys.

**Step 2 (Registration):** Each entity $i$ creates its own private/public ($k_i^-/k_i^+$) key pair and registers its public key with the RTA. For each registration entity, the RTA generates a unique ID and binds the entity's public key to its ID. If the registration entity is an RSU, the RTA stores the binding $< ID, public\ key >$ inside the RSU-PFD. On the other hand, if the registration entity is a vehicle, the RTA stores the entry inside the VPFD. In addition to storing the bindings $< ID, public\ key >$, the VPFD also stores the RTA's signature that is created from the vehicle's ID and its certified public key of each entry. This is to prevent a malicious RSU from issuing a signature on a wrong public key.

**Step 3 (Dissemination):** After a successful registration, the RTA disseminates a read-only copy of the VPFD to each RSU and a read-only copy of the RSU-PFD to each vehicle. Both directories are updated whenever a new entity joins the network or when an entity's public key is revoked. The size of the VPFD is finite as a vehicle will be de-registered and removed from the directory upon reaching its end-of-life. The RSU-PFD rarely increases in size because the number of RSUs in the network is fixed. The benefits of disseminating up-to-date directories are threefold: (1) eliminates the need to download huge certificate revocation list, (2) reduces the time and complexity required to verify the certificates and (3) improve the timeliness and freshness of public keys. Next, the RTA issues the $N$ plane equations and the hash functions $(H_1, H_2, H_3)$ to each entity which are kept in the HSM for protection.

TABLE II
NOTATIONS AND DESCRIPTIONS

| Notations | Description |
| --- | --- |
| m × m × m | Dimensions of 3D space |
| $(i_i, j_i, k_i)$ | 3D location of vehicle or RSU $i$ |
| $x_p, y_p$ | $x$ and $y$ coordinates of point $p$ in affine representation |
| $c_a, c_b$ | Constants of a plane equation |
| $H_1, H_2, H_3$ | Secure hash functions along each axis in 3D space |
| $k_i^-$ | Private key of vehicle or RSU $i$ |
| $k_i^+$ | Public key of vehicle or RSU $i$ |
| $R_i, C_i$ | Cryptographic nonce and commitment value of vehicle or RSU $i$ |
| $(e_i, S_i)$ | Signature pair issued by vehicle or RSU $i$ |
| $\mathcal{V}$ and $\mathcal{R}$ | Represent vehicle and RSU entity. |
| $v_i$ | Hash value issued by vehicle or RSU $i$ as proof |
| $\{M\}_{k_i^+}$ | Message M is encrypted by public key of vehicle or RSU $i$ |
| $Sig_{k_i^-}(M)$ | Signature of message M signed by private key of vehicle or RSU $i$ |
| $LOCM_i$ | GPS location of vehicle or RSU $i$ |
| $N_i$ | Nonce value of vehicle or RSU $i$ used in key agreement module |
| $T_i$ | Timestamp of vehicle $i$ used in key agreement module |
| $K_{i,j}$ | Pairwise key between vehicle $i$ and RSU $j$ |
| $exp_{LB}, exp_{UB}$ | Lower bound (LB) and upper bound (UB) of the expiry time of derived keys |
| $Dist_{LB}, Dist_{UB}$ | Lower bound (LB) and upper bound (UB) of the distance that derived keys stay valid. |
| $v_{avg}$ | Average speed of vehicle in m/s |

*B. Public Key Request Module*

This module is installed in each entity. It is invoked when the vehicle or the RSU receives public key request messages from the other entities. It contains an authentication mechanism similar to the one in [9] to mitigate the effects of internal and external DoS attacks. However, we improve the efficiency of the scheme based on the ECC technology. The key idea is to make the requestors expends an equal amount of computing power when it requests for a public key from the RSU. This involves verifying an EC-Schnorr signature to prove their identity and authenticity. Figure 2 illustrates the process between a vehicle $\mathcal{V}$ and an RSU $\mathcal{R}$ during the execution of the public key request module.

**Step 1 and Step 2:** First, $\mathcal{V}$ and $\mathcal{R}$ select one random value, $c_i \in [1, n-1]$ and a 128 bits random nonce $R_i$ where $i$ stands for the identifier of the entity and computes $C_i$ according to the formula in step 2 of Figure 2. The $R_\mathcal{V}$ and $R_\mathcal{R}$ values are nonce values to prevent the replay attacks while the $C_\mathcal{V}$ and $C_\mathcal{R}$ are commitment values used to provide entity authenticity and as a knowledge proof that the prover has verified the digital signature. The nonce and commitment values are pre-computed off-line or during the idle state of the entity's processor to make the scheme efficient.

**Step 3:** Suppose $\mathcal{V}$ wants to communicate with another vehicle, it constructs a public key query message $M_\mathcal{V}$ containing its own ID and the ID of the target vehicle, e.g. $ID_3$ and sends it to $\mathcal{R}$. This message also contains the $\mathcal{V}$'s random nonce value $R_\mathcal{V}$.

**Step 4:** Upon receiving $(M_\mathcal{V}, R_\mathcal{V})$ from $\mathcal{V}$, $\mathcal{R}$ generates a signature pair $(e_\mathcal{R}, S_\mathcal{R})$ consisting of a challenge, $e_\mathcal{R}$ and a response, $S_\mathcal{R}$ according to the formula given by step 4 in Figure 2 and sends the tuple $(e_\mathcal{R}, S_\mathcal{R}, R_\mathcal{R})$ to $\mathcal{V}$. The challenge $e_\mathcal{R}$ contains the hash of the message, including the random nonce values $(R_\mathcal{V}, R_\mathcal{R})$ and the $x_{C_\mathcal{R}}$ coordinate of the commitment value, $C_\mathcal{R}$ of $\mathcal{R}$ that is used as a proof to determine if $\mathcal{V}$ has verified the signature.

**Step 5:** When $\mathcal{V}$ receives $(e_\mathcal{R}, S_\mathcal{R}, R_\mathcal{R})$ from $\mathcal{R}$, it first retrieves the $\mathcal{R}$'s public key $k_\mathcal{R}^+$ from the RSU-PFD. Then, $\mathcal{V}$ re-computes the $\mathcal{R}$'s commitment value as $\bar{C}_\mathcal{R} = G \cdot S_\mathcal{R} + k_\mathcal{R}^+ \cdot e_\mathcal{R}$ using the values $S_\mathcal{R}$, $e_\mathcal{R}$ and $k_\mathcal{R}^+$. Next, $\mathcal{V}$ calculates its version of the challenge denoted as $\bar{e}_\mathcal{R}$ by taking the hash of the message, $M_\mathcal{V}$ concatenated with both the random nonce values $(R_\mathcal{V}, R_\mathcal{R})$ and the calculated $x_{\bar{C}_\mathcal{R}}$ coordinate of $\bar{C}_\mathcal{R}$ computed earlier. If the calculated hash value matches the received $e_\mathcal{R}$, it implies that $\mathcal{R}$ is authenticated successfully. If there is a mismatch, the session terminates at this stage. The proof of correctness is shown in (2) and (3).

$$\begin{aligned}\bar{C}_\mathcal{R} = G \cdot S_\mathcal{R} + k_\mathcal{R}^+ \cdot e_\mathcal{R} &= G \cdot (c_\mathcal{R} - b \cdot e_\mathcal{R}) + k_\mathcal{R}^+ \cdot e_\mathcal{R} \\ &= G \cdot c_\mathcal{R} - k_\mathcal{R}^+ \cdot e_\mathcal{R} + k_\mathcal{R}^+ \cdot e_\mathcal{R} \qquad (2) \\ &= c_\mathcal{R} \cdot G \end{aligned}$$

$$\bar{e}_\mathcal{R} = h(M_\mathcal{V}, R_\mathcal{V}, R_\mathcal{R}, x_{\bar{C}_\mathcal{R}}) \qquad (3)$$

If $\bar{e}_\mathcal{R} = e_\mathcal{R}$, the signature pair $(e_\mathcal{R}, s_\mathcal{R})$ from $\mathcal{R}$ passes the verification. Once the challenge $\bar{e}_\mathcal{R}$ is verified true, $\mathcal{V}$ prepares its own signature consisting of a challenge $e_\mathcal{V}$ and a response $S_\mathcal{V}$. At the same time, $\mathcal{V}$ computes a proof $v_\mathcal{V}$ which contains a hash of the two random nonce values and the $x_{\bar{C}_\mathcal{R}}$ coordinate of $\bar{C}_\mathcal{R}$. Message to $\mathcal{R}$ contains $(v_\mathcal{V}, e_\mathcal{V}, S_\mathcal{V})$.

**Step 6:** Before $\mathcal{R}$ proceeds to verify the signature pair $(e_\mathcal{V}, S_\mathcal{V})$ of $\mathcal{V}$. It first checks that the hash value $v_\mathcal{V}$ is correct by computing $h(R_\mathcal{V}, R_\mathcal{R}, x_{C_\mathcal{R}})$ using its own commitment value, $x_{C_\mathcal{R}}$ of $C_\mathcal{R}$ generated in step 2. The signature pair $(e_\mathcal{V}, S_\mathcal{V})$ needs to be verified only if the proof matches. If the proof provided by $\mathcal{V}$ is incorrect, it implies that $\mathcal{V}$ has not verified the signature in step 5. Therefore, $\mathcal{R}$ discards the message without verifying. If the hash value $v_\mathcal{V}$ is correct, $\mathcal{R}$ proceeds to locate the public key of $\mathcal{V}$ in the VPFD and performs the verification process to compute the commitment value of $\mathcal{V}$ as shown in step 6.

Lastly, $\mathcal{R}$ verifies if $e_\mathcal{V} = h(m, R_\mathcal{V}, R_\mathcal{R}, x_{C_\mathcal{V}}, C_\mathcal{R})$ matches the received $e_\mathcal{V}$. If $\overline{e_\mathcal{V}} = e_\mathcal{V}$, $\mathcal{R}$ issues the requested public key. Otherwise, $\mathcal{R}$ terminates at this step. The reply message to $\mathcal{V}$ contains the requested ID, the requested public key and the RTA's signature on the $< ID, public\ key >$ binding which are retrieved from the VPFD. The entire message is encrypted using $\mathcal{V}$'s public key to ensure confidentiality. The inclusion of the RTA's signature is to ensure the integrity and authenticity of the message. When $\mathcal{V}$ receives the reply message, it decrypts the message using its own public key $k_\mathcal{V}^+$ and verify the RTA's signature using $k_{RTA}^+$. If the signature is correct, it means that the requested public key issued by $\mathcal{R}$ is correct.

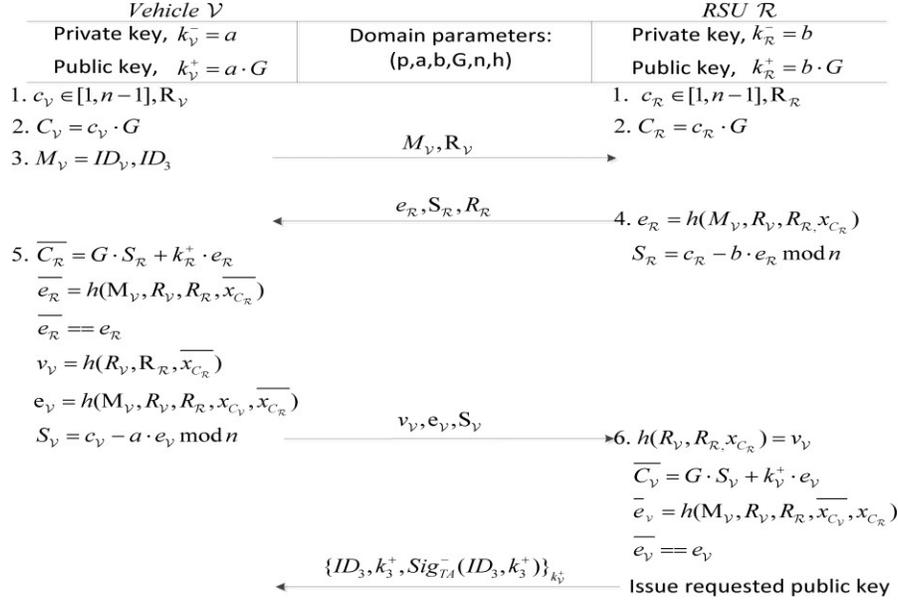

Fig. 2. Authentication process in the Public Key Request Module

### C. Key Agreement Module

Figure 3 describes the messaging protocol between two communicating entities during pairwise keys establishment. As the messaging protocol involves message exchanges, attackers can send invalid messages to deplete the resources of the recipients leading to DoS attacks. To solve this, we incorporate the authentication mechanism similar to the one used in the public key request module. The working principle is the same whereby both entities must prove to each other their intentions to derive the pairwise keys by committing some of their CPU resources. The only difference is that the exchanged messages contain additional information such as the randomly generated nonce $(N_\mathcal{R}, N_\mathcal{V})$ and the GPS information $(LOCM_\mathcal{R}, LOCM_\mathcal{V})$ of $\mathcal{R}$ and $\mathcal{V}$ respectively which is necessary for deriving the pairwise keys.

Once the GPS locations and the random nonce values $(LOCM_\mathcal{V}, LOCM_\mathcal{R}, N_\mathcal{V}, N_\mathcal{R})$ are known, $\mathcal{V}$ and $\mathcal{R}$ transit into the key derivation module to derive the pairwise keys. In this case, $\mathcal{V}$ can use any of the derived pairwise keys denoted as $K_{\mathcal{V},\mathcal{R}}$ to encrypt the messages via a symmetric algorithm. HMAC is appended to the message to prove authenticity and integrity. The timestamp is also included in the message to prevent replay attacks. In order for $\mathcal{R}$ to know which pairwise key is used to decrypt the actual message, the pairwise key is indexed by $keyID$ and sent to $\mathcal{R}$. The format of the message sent by vehicle $\mathcal{V}$ to RSU $\mathcal{R}$ is given as: $keyID, \{Message, T_\mathcal{V}, HMAC(K_{\mathcal{V},\mathcal{R}}||Message||T_\mathcal{V})\}_{K_{\mathcal{V},\mathcal{R}}}$. When $\mathcal{R}$ receives the message from $\mathcal{V}$, it uses the corresponding pairwise key indexed by the key ID to decrypt the actual message. Our pairwise key agreement protocol can also be extended to support group communications in a V2V context with the help of an RSU. After the RSU has derived the pairwise keys, it can behave as a mediator to authenticate the other vehicles and distribute the set of common keys to each of the authenticated vehicles that the requesting vehicle wants to communicate with. Once the group vehicles have received the common keys, they can start the secure communication within the group without further interactions with the RSU. Moreover, our scheme can also be applied to the pairwise V2V communications. However, the initiating vehicle has to contact the RSU first to request for the recipient's public key because the vehicles do not have the VPFD repository. Thereafter, the rest of the procedure in V2V follows the messaging protocol in Figure 3.

### D. Key Derivation Module

This module leverages on the 3D matrix key approach to generate the pairwise keys. However, instead of preloading the keys onto the vehicles and the RSUs, each entity derives the keys dynamically. We illustrate the procedure of obtaining the pairwise keys in a V2I communication where each entity has $N$ plane equations as defined in equation (1).

Both entities have to solve the $N$ plane equations in groups of 3 to determine the solutions. For example, $\mathcal{V}$ inputs its own GPS location $(i_\mathcal{V}, j_\mathcal{V}, k_\mathcal{V})$ into the first plane equation and substitute the GPS location $(i_\mathcal{R}, j_\mathcal{R}, k_\mathcal{R})$ of $\mathcal{R}$ into the second plane equation in (1). For the third plane equation, $\mathcal{V}$ uses its own randomly generated nonce value $N_\mathcal{V}$ (Message 2 of the key agreement protocol) as the first input for $i_i$ and the second nonce value $N_\mathcal{R}$ generated by $\mathcal{R}$ (Message 3 of the key agreement protocol) as the second input for $j_i$. The last input $k_i$ of the third equation can be found by either hashing the first and the second nonce values or XOR-ing the two nonce values together. The random nonce values guarantee the uniqueness of the derived keys. It ensures that another vehicle with the same GPS coordinates communicating with the same RSU will not derive the same keys as the vehicles. With 3 sets of inputs to substitute into the $N$ plane equations, there are $N(N-1)(N-2)$ equation groups to solve which generate $N(N-1)(N-2)$ unique solutions. The set of solutions is described by $S = \{(x, y, z)_1, (x, y, z)_2, \cdots, (x, y, z)_d\}$ where $(x, y, z)$ denotes a solution point and $d$ denotes the number of solutions in the set. To derive the pairwise key at a particular solution point $(x, y, z)$, we define equation (4) where the repeated hash results of the concatenation of the two random nonce values and the key requestor's ID is XORed together. In this case, the number of hashing operations is determined by the value of $x, y$ and $z$ in the solution point.

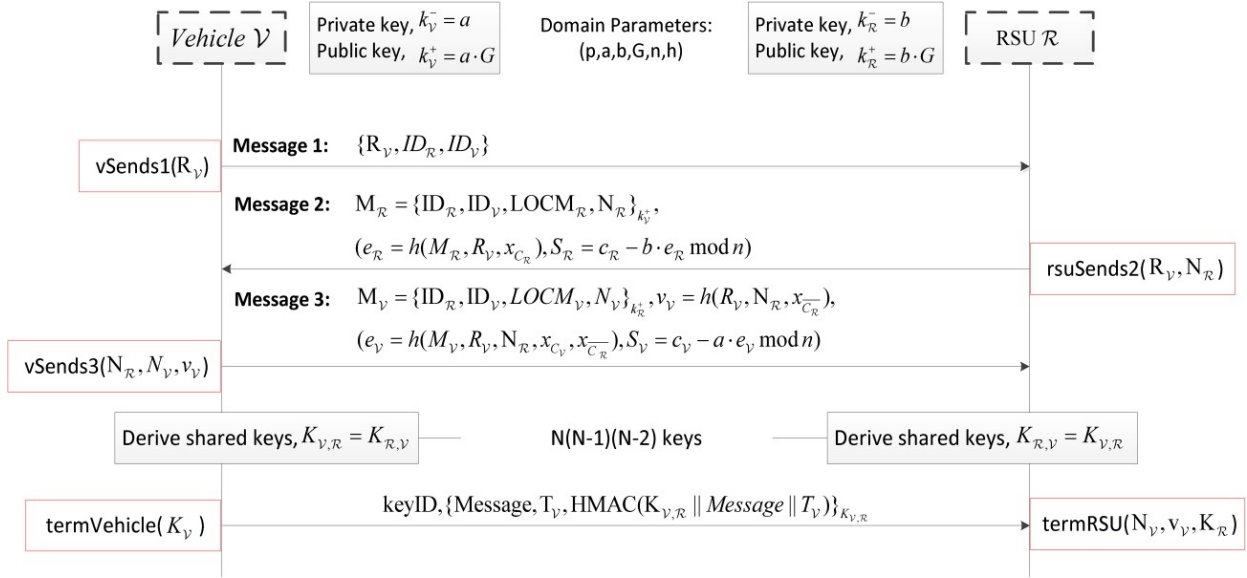

Fig. 3. Exchange of keying materials for establishing pairwise key in V2I

$$K_{V,R}(x, y, z) = H_1^x(N_V \| N_R \| ID_V) \oplus H_2^y(N_V \| N_R \| ID_V) \oplus H_3^z(N_V \| N_R \| ID_V) \quad (4)$$

where $H^\beta(\cdot)$ denotes that the hashing function $H(\cdot)$ is repeated for $\beta$ times. Similarly, $R$ solves $N(N-1)(N-2)$ equation groups and performs the same operations as described in (4) to compute the pairwise keys. It can be shown that for any key, this condition holds: $K_{V,R} = K_{R,V}$. Once the pairwise keys are obtained, we can use any key from the $N(N-1)(N-2)$ key pool for symmetric encryption or decryption. We can also choose a subset of $\delta$ keys from the key pool to compute a composite key using the recursive XOR operation as shown in (5) where $(x, y, z)$ is an element of the solution set $S$ and contains $d$ solutions.

$$K_{composite,V,R} = \oplus K_{V,R}(x, y, z) \quad where \; (x, y, z) \in S \; and \; 2 \leq \delta \leq d \quad (5)$$

*E. Revocation Module*

In our scheme, both the public key request module and the key agreement module rely on the EC-Schnorr signature for mutual authentication. If a vehicle fails the authentication, the RSU will send a message to the RTA to revoke that vehicle's public key. RTA will update the VPFD and disseminate it to all the RSUs in the region via a secure network in real time. It implies that future communications with that vehicle are disabled unless it re-registers itself with the RTA again to receive another set of public/private key pairs. After that, the new updated VPFD containing the vehicle's new public key will be disseminated to all the RSUs.

On the lifetime of the pairwise keys, we assume that each vehicle and RSU sends a message between 100ms-300ms according to the WAVE standard [23] and [24]. Furthermore, suppose that a shared key can only be used for one message, the lower bound and upper bound of the expiry time of the derived keys can be determined using (6).

$$exp_{LB} = (0.1 * \# \; of \; common \; keys) \; in \; seconds$$
$$exp_{UB} = (0.3 * \# \; of \; common \; keys) \; in \; seconds \quad (6)$$

where $\# \; of \; common \; keys = N(N-1)(N-2)$ and $N$ is the number of plane equations. If $N = 10$, two communicating nodes get to keep 720 keys for the duration between 72 seconds and 216 seconds before they are destroyed. When the keys expire, the nodes need to re-initiate the key agreement module again. Using this information, we can further determine how many RSUs need to be informed about the derived keys in a V2I communication. The formula is given in (7).

$$Dist_{LB} = (V_{avg} * exp_{LB}) \; in \; meters$$
$$Dist_{UB} = (V_{avg} * exp_{UB}) \; in \; meters \quad (7)$$

where $V_{avg}$ is the average speed of the vehicle. Suppose the RSUs are 1km apart and the average speed of the vehicle is 100km/h on a highway, the number of RSUs to inform is between 2 and 6. It means that a vehicle does not have to establish shared keys each time it passes an RSU which can greatly reduce the rekeying overhead.

## V. SECURITY ANALYSIS OF SA-KMP

We analyze the SA-KMP scheme to prove that it is secured against classical attacks such as DoS, eavesdropping, data modification, replay, GPS spoofing, node impersonation, repudiation, and collusion attacks. For more details on the definitions of these attacks, we refer the readers to [2] and [8]. We further provide a formal security validation of our scheme using the Proverif tool [29].

### A. Classical Attacks

*DoS Attack* – By the proposed SA-KMP scheme, DoS attacks are mitigated because the vehicle has to verify the signature pair of the RSU to extract the commitment value for generating the hash proof as described in Figure 2 and Figure 3. The vehicle may try to forge a valid hash proof containing the commitment value. However, it is not possible due to the property of the one-way hash function. To further illustrate the effectiveness of our scheme, we calculate the authentication delay when the vehicles flood many signatures to the RSU and compare the results to the PKR scheme [18] and the Wasef's scheme [27]. The authentication delay is defined as the time incurred by the RSU to verify the signatures from the vehicles. We use the ECDSA verification times in Table V and the execution time of the HMAC function as 0.008 ms based on 100000 simulation runs to calculate the authentication delay.

Figure 4 shows the authentication delay versus the number of valid messages for two cases: 1) no DoS and 2) DoS attacks. We observe that when the PKR scheme is deployed, the introduction of invalid messages (10% and 30% of the number of valid messages) increases the authentication delay substantially because the PKR scheme has no defense mechanism to mitigate the DoS attacks. Consequently, the PKR scheme has to verify all the signatures even if they are invalid. On the other hand, the Wasef's scheme appends an HMAC to all the outgoing messages when the number of invalid signatures to the number of valid signatures exceeds a threshold. Therefore, the authentication delay is slightly higher than the PKR scheme where it consists of the sum of all the ECDSA verification times and the HMAC verification times. In contrast, the authentication delay for the SA-KMP scheme is significantly lower than that of the PKR scheme and Wasef's scheme under no DoS attack. Furthermore, in the left inset figure of Figure 4, the performance of the SA-KMP scheme when subjected to 10% and 30% invalid messages, has little effect on the authentication delay and is also significantly lower than the PKR scheme and the Wasef's scheme under DoS attacks. The SA-KMP scheme is efficient because the RSU only has to perform a light step to validate the hash proof containing its commitment value. If the proof is incorrect, RSU will skip the signature verification. The authentication delay of our scheme, in this case, is the sum of the signature generation time and the hashing time. A lower authentication delay indicates a higher availability to service the requests of other vehicles. Based on these observations, we conclude that SA-KMP does not introduce additional delay and can mitigate DoS attackers effectively even though security measures are introduced.

*Eavesdropping Attack* – The proposed SA-KMP scheme is secure against the eavesdropping attacks because all the messages are encrypted to ensure confidentiality. By the 3D matrix based scheme, if $m$ is selected to be a prime and the rank of the coefficient matrix is of full rank, then two communicating entities have in common $N(N-1)(N-2)$ distinct keys where $N$ is the number of plane equations. It implies that the probability of guessing the correct derived key is $1/N(N-1)(N-2)$ which is low if $N$ is large. Moreover, if the SHA-1 encryption is used, each derived key is 160 bits in length. The computational complexity to find the correct key is therefore $O(2^L)$ where $L =$ the length of the key in bits. Furthermore, two communicating entities can derive a composite session key using some of the derived keys to increase the difficulty of cracking the keys. For example, if $N = 10$, then there are 720 derived keys between two communicating entities. If any 50 keys are selected to compose the composite key, there will be at least $\binom{720}{50}$ possible combinations of the composite keys.

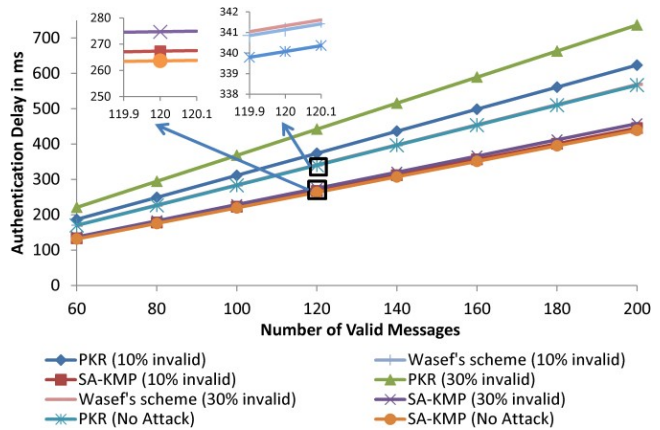

Fig. 4. Authentication delay under different scenarios

*Data Modification Attack* – By the proposed SA-KMP scheme, all the messages in the public key request module and the key agreement module are digitally signed to detect data modifications. Furthermore, in the key agreement module, when the shared keys have been established between two entities or among a group, an HMAC using one of the derived keys is appended to the message to check for the integrity and authenticity of the message.

*Replay Attack* –By the SA-KMP scheme, random nonce values are used in the public key request module including the key agreement module to prevent the replay attacks. The nonce is at least 128 bits long which implies that the probability of getting the same random nonce for two different communication sessions is equal to $1/2^{128}$. Also, all the messages are time-stamped to ensure the freshness. If the message is not received within a tolerable period, HMAC verification will fail, and the packet will be discarded.

*GPS Spoofing Attack* - In the SA-KMP, the external attackers are not able to intercept the message and modify the GPS coordinates because all the messages are encrypted using asymmetric cryptography during the key agreement phase. Furthermore, the private key is protected and stored in the HSM which is programmed to self-erase when any physical tampering is detected. On the other hand, an internal attacker may misbehave and modify the GPS coordinates in the LOCM message. This form of attack can be prevented because the vehicles and the RSUs are equipped with a GPS receiver to perform cross verification on the received GPS coordinates. It can also be avoided by employing the verifiable multilateration method [28] to verify the vehicles' positions. Hence, a GPS spoofing attack cannot succeed.

*Node Impersonation Attack* - By the proposed SA-KMP scheme, the vehicles, and the RSUs are given a public key repository where it contains the certified node ID and public key bindings issued by the RTA during the registration phase. This information is used to authenticate each other mutually in the key agreement step. Therefore, impersonation cannot succeed. The position of the nodes is also verified in the process which makes the impersonation even harder. Using shared keys for the symmetric encryption can also provide some level of authentication as the shared keys are only known to a group of vehicles.

*Repudiation Attack* – By the proposed SA-KMP scheme, non-repudiation is achieved by employing digital signature in the public key request module and the key agreement module as well as requesting each vehicle to sign the key index of the shared key.

*Collusion Attack* – By this attack, two or more vehicles may collude to capture the nodes to compromise the session keys or on a larger scale, to capture the whole key space. For this reason, the matrix based key [10], [11] and the probabilistic key sharing [12] schemes are required to change the keys periodically which increase the communication overhead. In the design of the proposed SA-KMP scheme, the keys are not preloaded to increase the resilience to node capture attacks. Instead, the key derivation module generates the keys on demand based on the current inputs to the plane equations. Even if a particular group of network entities is compromised, it only affects the communication in that particular region without jeopardizing the whole network. Moreover, the public key request module of the proposed SA-KMP scheme is resilient against the collusion attacks. Interested readers may refer to [9] for more details.

### B. Formal Verification by Proverif

Proverif [29] is a software designed based on the Dolev and Yao attack model [21] to analyze the security of cryptographic protocols. It can verify whether a protocol satisfies secrecy, correspondence assertions and observational equivalence properties. In this paper, the secrecy and correspondence assertions of the key agreement module in Figure 3 are validated. To test the correspondence property, we define five events in the sub-processes of the vehicle and the RSU in Proverif. These five events which are also indicated in Figure 3 are: (1) Event vSends1($R_V$), (2) Event rsuSends2($R_V$, $N_R$), (3) Event vSends3($N_R$, $N_V$, $v_V$), (4) Event termRSU($N_V$, $v_V$, $key_R$) and (5) Event termVehicle($key_V$). These events create an ordered association between messages in the protocol to prove authentication property i.e. if an event has been executed, it implies another event has been previously executed. To verify the authenticity of the RSU, the following query syntax is used:

**query rand:nonce, rand1:nonce, rand2:nonce, rand3:bitstring, k1:key;**
**inj-event(termVehicle(k1))==>(inj-event(termRSU (rand2,rand3,k1))==>(inj-event(vSends3(rand1,rand2, rand3))==>(inj-event(rsuSends2(rand,rand1))&&inj-event (vSends1 (rand))))).**

The above relationship implies that each termination of the vehicle has to be preceded by an instance of the following events: termRSU(), vSends3(), rsuSends2() and vSends1(). If the query result is true, it means that the RSU is properly authenticated. Otherwise, it is not. Similarly, we use the following query format below to test that the RSU completes the protocol with the target vehicle only when events vSends3, rsuSends2, and vSends1 are executed.

**query rand:nonce, rand1:nonce, rand2:nonce, rand3:bitstring, k1:key;**
**inj-event(termRSU(rand2,rand3,k1)) ==> (inj-event (vSends3(rand1,rand2,rand3)) ==> (inj-event (rsuSends2 (rand,rand1)) && inj-event(vSends1(rand)))).**

For both correspondence queries, the injective correspondence is used to capture the one-to-one relationship between the number of protocol runs performed by each participant. When the injective query is violated, it means that the protocol is subject to a replay attack and does not fulfill the authenticity property. Next, we verify the secrecy of the shared key by declaring **out(c, senc(secretmessage, derived_key_rsu))** in the sub-process of the RSU. It means outputting a free name **secretmessage** encrypted using a key derived by the RSU onto the channel c. Then, we test the secrecy of the free name using the following query: **query attacker(secretmessage)**. The free name **secretmessage** is secret if and only if the derived key, denoted by **derived_key_rsu** is secret. The results of executing the key agreement protocol by Proverif are shown in Figure 5. Line 1-2 indicates that the attacker is unable to determine the derived key shared between the RSU and the vehicle. Hence, the secrecy of the derived key is preserved. Moreover, the results given in line 3-7 indicate that the injective authentication of the vehicle to the RSU and vice versa hold which means that the RSU and the vehicle participating in the protocol are both authenticated. In addition, it proves that the replay and the impersonation attacks are impossible to succeed by the attacker. As such, we conclude that our key agreement protocol is secure against an active attacker under the Dolev and Yao model.

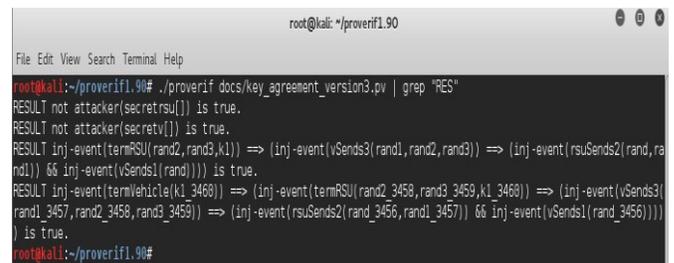

Fig. 5. Proverif verification results

### VI. PERFORMANCE ANALYSIS OF THE SA-KMP

In this section, we evaluate the performance of the SA-KMP scheme in terms of the transmission and storage overhead, latency, scalability, key generation time and computational complexity.

## A. Transmission Overhead

We analyze the transmission overhead of the SA-KMP scheme and compare its performance with the certificate-based scheme in [5] and the PKR scheme in [18]. The comparison is made by evaluating the transmission overhead when a vehicle sends a packet to an RSU or vice versa. By the certificate-based scheme, the size of the certificate for a vehicle is 126 bytes and the size of the ECDSA signature that is created by a vehicle based on a key length of 224 bits is 56 bytes. Therefore, the total transmission overhead of a message is $126 + 56\ bytes = 182\ bytes$ as shown in Table III. By the SA-KMP scheme and the PKR scheme, the vehicles and the RSUs are given a copy of the directory which contains the public keys that have previously been verified by the RTA. Therefore, there is no need to attach the certificate in the message transmission which reduces the total transmission overhead to 56 bytes as shown in Table III. We further illustrate the transmission overhead improvement in Figure 6 by plotting the transmission overhead as a function of the number of messages received by an RSU in 30 seconds. It is observed that both the SA-KMP and the PKR schemes have the lowest transmission overhead compared to that of the certificate based scheme even when the number of messages received at the RSU increases. The transmission overhead of the SA-KMP is 30.8% that of the certificate based scheme which means that our scheme saves about 69.2% of the communication bandwidth without transmitting certificates.

## B. Storage Overhead

We estimate the storage required by each RSU and vehicle to maintain the VPFD and RSU-PFD repositories. According to [5], the public key size of an RSU and a vehicle is 33 bytes and 29 bytes respectively, and the size of the RTA's ECDSA signature is 64 bytes. If 4 bytes are used to store the vehicle ID and assuming there are 1 million vehicles in the network, the size of the VPFD held by each RSU would be 1 million x $(29 + 4 + 64)$ bytes = 97 Mbytes as shown in Table IV, which is higher than 33 Mbytes required by the PKR scheme. This additional storage requirement is a trade-off for the enhanced security against the collusion attacks. On the other hand, if there are 100,000 RSUs deployed in the network, the size of the RSU-PFD which is maintained by each vehicle is equal to 100,000 x $(33 + 4)$ bytes = 3.7Mbytes as shown in Table IV. Assuming the storage capacity of each vehicle and each RSU is 256M bytes, the storage costs of 97M bytes and 3.7M bytes take up only about 37.9% and 1.45% of the storage space in each RSU and vehicle, respectively which is reasonable. In contrast, the certificate based PKI scheme requires each vehicle to store a set of certificates for the PKI support as suggested by [22] and [25]. However, the number of certificates to store is not specified and can vary according to various application requirements. For this reason, we express the storage overhead as a function of the number of certificates in Table IV. We note here that even if the storage requirement is lower than our scheme, the certificate based PKI scheme suffers from high latency due to the cumbersome certificate management.

## C. Latency Analysis

Next, we analyze the latency of our scheme and compare the results to the certificate based PKI scheme [5] and the PKR scheme [18]. We define the latency as the amount of time a packet takes to travel from the source to the destination including the time taken to acquire the public key for verifying the message. In our analysis, we omit the transmission delay and the queuing delay and evaluate the latency as the sum of the propagation delay and the processing delay. Figure 7 illustrates the timing diagrams of the various schemes whereby the vertical arrows represent the processing delay while the diagonal or horizontal arrows represent the propagation delay.

According to Figure 7(i), a vehicle has to send a message to an RSU to request for the public key of the target vehicle. To mitigate DoS attacks during the public key request, both vehicle and RSU exchange EC-Schnorr signatures which involve verification of proof. Thus, the main bulk of the processing delay in a V2V communication consists mainly of signature generation and verification times including the generation and verification of proof from both sides. On the other hand, the processing delay in a V2I communication is much simpler as the vehicle and RSU has a copy of the RSU-PFD and VPFD respectively which eliminates the need to request a public key. Hence, the processing delay is $T_{search} + T_{verify,ECDSA} + T_{sign,ECDSA}$ while the propagation delay is $T_{tx,signed\ message\ without\ cert.}$. By the certificate based PKI as shown in Figure 7(ii), the main delay is due to the downloading of the CRL and the verification of the certificates. Lastly, in the PKR scheme as illustrated in Figure 7(iii), the latency consists of the delay in acquiring the public key from the RSU which involves generation and verification of an ECDSA signature. To evaluate the latency of each scheme numerically, we first implement the EC-Schnorr and the ECDSA signature schemes in C language using the Openssl package. Simulation has been carried out for 100000 times on a 32 bit Debian Linux operating system with 2 GB RAM running on an Intel Core i7-2620M@2.70Ghz workstation. The signing and verification times of both signature schemes is averaged and presented in Table V. To calculate the propagation delay, we assume the followings according to the parameters given in [18]: 1) the size of an unsigned message = 69 bytes, 2) the size of a signed message = the size of an unsigned message (69 bytes) + the size of an ECDSA signature (56 bytes) = 125 bytes and 3) the size of a signed message including the certificate = size of a signed message (125 bytes) + size of a certificate (126 bytes) = 251 bytes. The rest of the parameters as presented in Table VI are set according to the simulation results presented in [18].

TABLE III
COMPARISON OF TRANSMISSION OVERHEAD

| Scheme | Sending a message |
|---|---|
| Certificate based PKI [5] | 182 bytes |
| SA-KMP | 56 bytes |
| PKR [18] | 56 bytes |

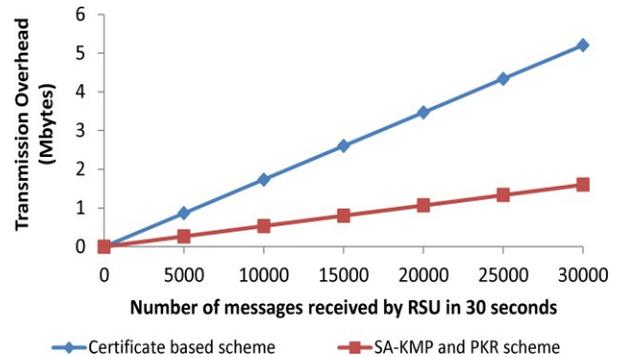

Fig. 6. Transmission overhead as a function of messages.

TABLE IV
COMPARISON OF STORAGE OVERHEAD

| Scheme \ Repository | RSU-PFD | VPFD |
|---|---|---|
| SA-KMP | 3.7M bytes | 97M bytes |
| PKR [18] | - | 33M bytes |
| Certificate based PKI [5] | 126bytes per certificate x no. of certificates | |

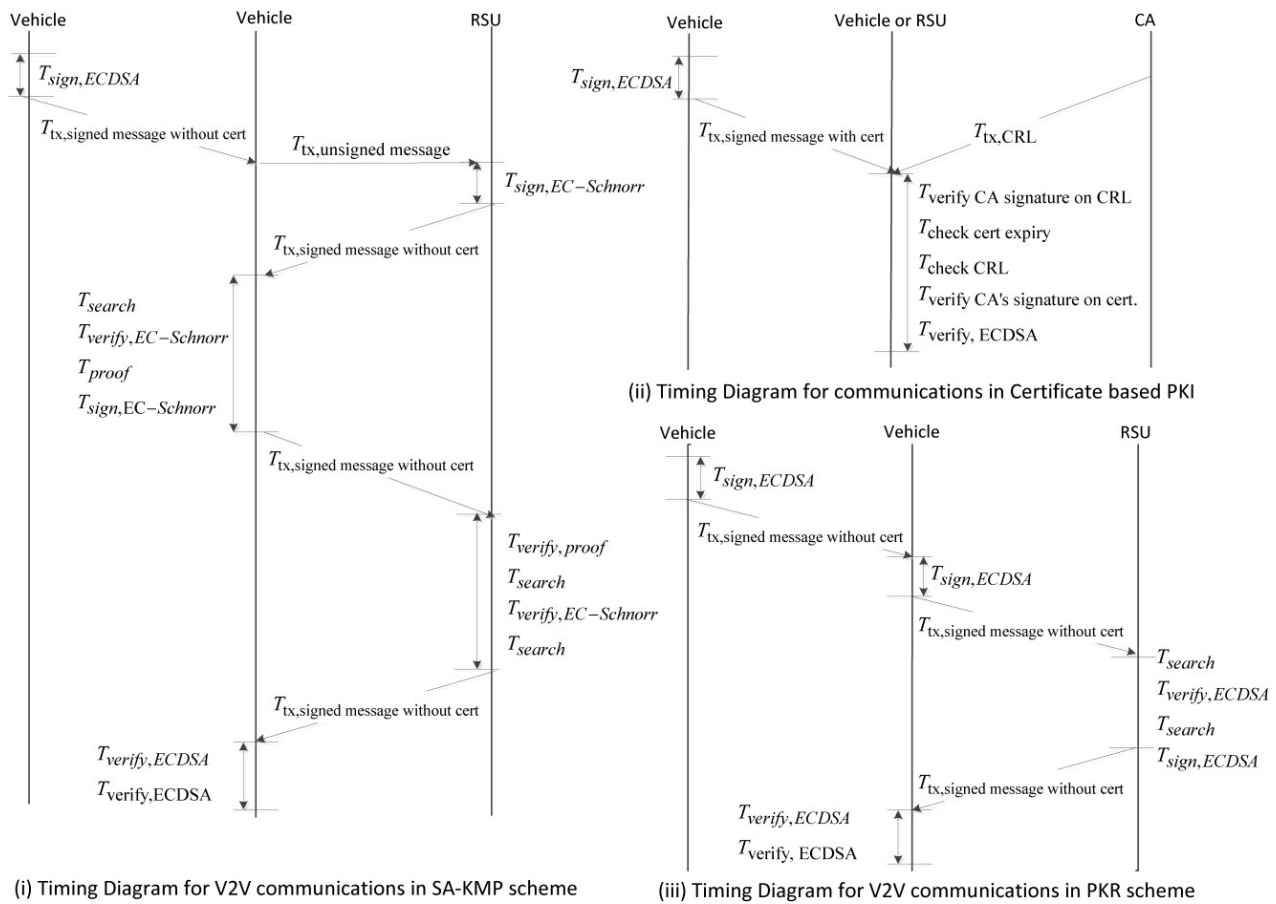

Fig. 7. Latency of SA-KMP, Certificate based PKI and PKR Scheme

With the help of Figure 7 and the values in Table V and Table VI, we tabulate the total latency of the three schemes in Table VII. Results show that the latency involved in the exchange of certificates and downloading of CRL by the certificate based PKI is several orders of the magnitude higher than that incurred in a V2V and V2I communication by the SA-KMP scheme. Moreover, the V2V communication latency by the SA-KMP scheme is lower than that by the PKR scheme even though an authentication is introduced against DoS attacks. Our scheme is more superior because 1) no certificates are sent in the message transmission and 2) the EC-Schnorr signing operation is faster than the ECDSA as it does not involve modular inversions. Our scheme is also able to support V2I communication with a low network latency of 4.488ms which has not been addressed by the PKR scheme. These results show that the proposed SA-KMP scheme cannot introduce a very long delay time in acquiring public keys.

TABLE V
SIGNATURE SIGNING AND VERIFICATION TIMES

| Algorithm<br>Key length | EC-Schnorr<br>256 | ECDSA<br>256 |
|---|---|---|
| Signing (ms) | 0.154 | 1.486 |
| Verification (ms) | 1.888 | 2.834 |
| Generate /Verify Proof (ms) | 0.155 | - |

TABLE VI
OTHER EVALUATION PARAMETERS

| Parameters | Time |
|---|---|
| $T_{search}$ | 0.000859 ms |
| $T_{check\ cert\ expiry}$ | 0 ms |
| $T_{check\ CRL}$ | 97.2975 ms |
| $CRL_{size}$ | 0.9 Mbytes |
| Data rate | 6Mbps |

TABLE VII
NETWORK LATENCY OF DIFFERENT SCHEMES

| Schemes in V2V and V2I/I2V | Network latency (ms) |
|---|---|
| SA-KMP (V2V only) | 12.309 |
| SA-KMP (V2I/I2V) | 4.488 |
| Certificate based with CRL downloading | 1365.911 |
| Certificate based without CRL downloading | 117.097 |
| PKR (V2V only) | 13.462 |

### D. Scalability

In this section, we analyze the scalability of our scheme for an increasing number of users and compare our results to the certificate based PKI scheme [5] and the PKR scheme [18]. The scalability is evaluated in terms of three aspects: the transmission overhead, storage overhead, and the latency. In assessing the transmission overhead of our scheme, we consider two cases, namely the V2V and V2I communication. In the case of a V2V communication, a vehicle has to perform two rounds of authentication before the actual communication takes place. The first authentication is to retrieve the public key of the target vehicle while the second set of authentication is to exchange keying materials for the establishment of pairwise keys. Each round of the authentication process contains an EC-Schnorr signature for mitigating DoS attacks. On the other hand, in a V2I communication, only one round of authentication is needed since the vehicle and the RSU have each other public key. Suppose the EC-Schnorr signature is 56 bytes, the total transmission overhead generated by a vehicle in a V2V communication will be 112 bytes while the transmission overhead in a V2I communication is 56 bytes. In the PKI scheme, the transmission overhead is 182 bytes as given in Table III whereas, the transmission overhead of the PKR scheme is 56 bytes. In terms of storage overhead, we analyze the storage requirements for each vehicle and RSU in our scheme separately. We assume the number of RSU in the network remains constant at 1000 regardless of the increase in the vehicle population. In the certificate based PKI scheme, we assume that the storage overhead is due to the need to download and store the CRL list. Thus, we estimate the size of the CRL given in Table VIII based on a revocation rate of 10% and assuming that the CRL header size is 50 bytes, the RTA's signature on the CRL is 64 bytes and each revoked certificate is 9 bytes.

Figure 8 shows a composite graph where the primary axis denotes the transmission overhead in logarithmic values while the secondary axis, also in logarithm scale, denotes the storage overhead requirements. From Figure 8, we observe that the transmission overhead of our scheme in a V2I communication is the same as the PKR scheme. But, in the case of a V2V communication, our scheme incurs a higher transmission overhead. This is due to the authentication mechanism introduced in the key agreement module to mitigate the flooding of invalid signatures. Nevertheless, the results suggest that the transmission overhead of our scheme is still lower than the classical PKI scheme. Suppose the typical data rate of a vehicular network is 27Mbps [23] and the transmission overhead in a V2V and V2I communication is 112 bytes and 56 bytes respectively, it would mean that SA-KMP is able to support between 30,000 and 60,000 users concurrently in a V2V and V2I communication respectively. In contrast, the certificate based PKI scheme could only support up to 18,500 users since the transmission overhead is high at 182 bytes due to the sending of a certificate and an ECDSA signature for each communication. Even though the PKR scheme can support about 60,000 users which is similar to the SA-KMP in a V2I communication, we emphasize that the PKR scheme does not have protection against DoS attacks which is the main benefit of our scheme. With regard to the storage overhead, it is evident that VPFD in our scheme is higher than the PKR scheme as we are storing the RTA's signature in the repository. However, the RSU-PFD which is maintained by each vehicle remains constant throughout because the RSU population rarely increase in the network. Suppose the storage space to store an entry corresponding to a vehicle and RSU is 97 bytes and 37 bytes respectively (discussed in Section VI-B), the total storage space required to store 1 million vehicles and 1000 RSUs by each RSU and vehicle will be about 97 Mbytes and 0.037 Mbytes respectively. This storage requirement is negligible given that modern-day OBU has a storage capacity of 8GB [26]. This observation supports our claim that SA-KMP is highly scalable in terms of increasing users. Even though the certificate-based scheme has the lowest storage overhead, the main disadvantage of using certificates is that the latency is high due to the downloading of CRL which we analyze next.

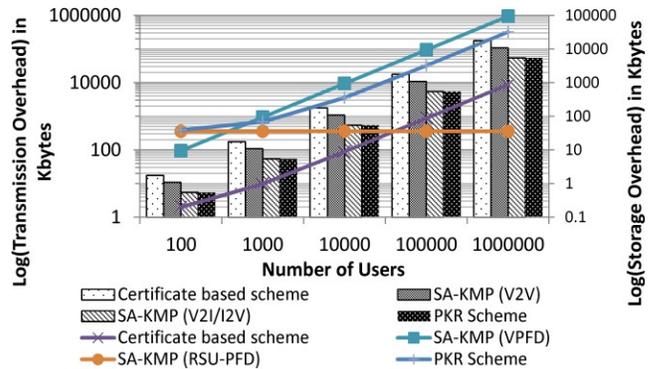

Fig. 8. Scalability comparison in terms of transmission overhead and storage overhead

TABLE VIII
Various Parameters For Calculating The Latency

| Number of Users | Number of revoked certificates | $CRL_{size}$ (Bytes) | Certificate based PKI $T_{check\ CRL}$ (ms) | SA-KMP $T_{search}$ (ms) |
|---|---|---|---|---|
| 100 | 10 | 204 | 0.023 | 0.000667 |
| 1000 | 100 | 1014 | 0.071 | 0.00135 |
| 10000 | 1000 | 9114 | 0.691 | 0.001663 |
| 100000 | 10000 | 90114 | 7.910 | 0.003196 |
| 1000000 | 100000 | 900114 | 125.987 | 0.008354 |

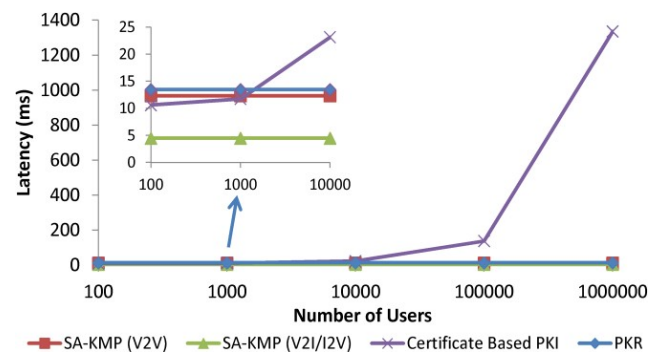

Fig. 9. Scalability in terms of latency experienced by a user

To evaluate the latency of our scheme, we develop a C program to measure the time taken to search for a public key in the VPFD. We further develop another C program to estimate the time required to check the validity of the certificate against the CRL of the certificate based PKI scheme. The search operations by both programs are evaluated for a different number of users ranging from 100 to 100000 and the average timing is taken over 100 simulation runs. Table VIII shows the simulation results for different parameters under considerations. It is obvious that the searching time by the PKR scheme is the same as that of the SA-KMP scheme. Using the values in Table V and Table VI, we calculate the latency of each scheme by evaluating the propagation delay and the processing delay as per Figure 7. Figure 9 shows the latency performance of the three schemes for a different number of users in the network. According to Figure 9, the latency to deploy the certificate based PKI scheme increases sharply when the number of users is beyond 10,000. This is due to the large CRL which results in a longer time to download and to search for a certificate. On the other hand, the latency in a V2V and V2I communication based on our scheme remains constant at around 12.5ms and 5ms irrespective of the number of users in the network. Furthermore, according to the left inset figure in Figure 9, the latency of the SA-KMP scheme is lower than that of the PKR scheme even though an authentication mechanism is introduced to mitigate the DoS attacks. These results conclude that our scheme is highly scalable to cope with an increasing number of users.

*E. Key Generation Time*

We develop a C program based on the key derivation module described in our scheme to calculate the time required to compute a pairwise key. The program is developed to solve a system of plane equations and performing hashing operations to derive the keys. First, we investigate the effects of varying the network size on the key generation time. After that, we study the relationship between the key generation time and the number of plane equations.

In the first experiment, we fix the number of plane equations $N$ as 3 and we vary the modulus $m$ operator in the equation (4) to be a prime number that approximates the network size between 0.5km and 10km. After that, we run the simulation 1000 times to measure the average key generation time and compare the results of our scheme to the ECDH protocol [17] and the DH protocol [16]. Results in Figure 10 show that the average key generation time of the SA-KMP scheme is much lower than the DH protocol across the whole range of network sizes. Our scheme also outperforms the ECDH-256 and ECDH-224 protocols when the network size is smaller than 9km and 8km respectively. Furthermore, it can be seen in Figure 10 that the key generation time increases linearly with the network size. This is because the common keys are derived based on hashing the equation (4) a number of times according to the solution point. As the network size increases, it requires more hashing operations to determine the keys. To make our scheme more scalable in terms of the network size, we can reduce the number of hashing operations by taking the modulus ($num$) of the solution points. The modulus $num$ can either be a fixed parameter pre-determined by the RTA during the registration phase or it can be a parameter determined by the key requestor during the key agreement phase. The extended version of the SA-KMP scheme with reduced hashing operations is simulated using modulus $num = 100$ and the result is represented by the line with the green triangular markers in Figure 10. With reference to the secondary axis on Figure 10, the average key generation time for the extended version of the SA-KMP scheme is reduced and it remains constant at about 0.025ms throughout the various network sizes.

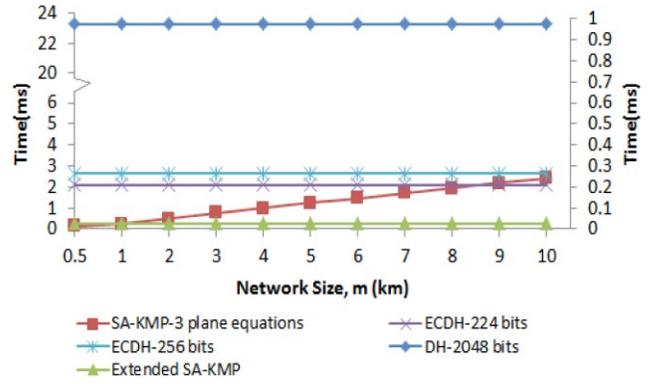

Fig. 10. Key generation time as a function of network size

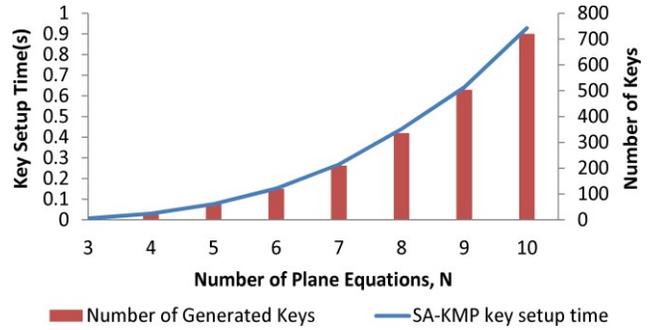

Fig. 11. Key generation time as a function of plane equations

Next, we study the effects of increasing the number of plane equations $N$ on the key generation time. In this experiment, the network size is fixed at 5km with modulus $m = 4999$ and the number of plane equations is varied from 3 to 10. The simulation is repeated for 1000 times to compute the average key generation time. Figure 11 shows that the key generation time increases exponentially with the number of plane equations. This is because there are $N(N-1)(N-2)$ groups of equations to solve and $N(N-1)(N-2)$ keys being generated for $N$ plane equations. Although simulation result shows that it only takes about 0.93 seconds to generate 720 keys with 10 plane equations, the key generation time is going to scale up exponentially with respect to $N$ which renders our scheme time-inefficient. To address this issue, we can randomly solve any combinations out of $\binom{N}{3}$ where $N$ is the number of plane equations given. For example, if $N = 4$ and assuming all the plane equations issued by the RTA are indexed, $\binom{N}{3}$ results in the following combinations where *#number* refers to the ID of the given plane equation: {#1, #2, #3}, {#1, #2, #4}, {#1, #3, #4}, {#2, #3, #4}. The key requestor can choose any sets out of the 4 combinations above to derive the common keys. If the key requestor chooses 3 sets from the possible combinations to solve, there will be 18 keys instead of 24. In this way, key generation time can be reduced. This makes our scheme highly configurable and flexible.

*F. Computational Complexity*

Next, we evaluate the computational complexity of the SA-KMP scheme by counting the number of CPU cycles it takes to execute in both V2V and V2I communication scenarios. In the V2V communication, both communicating parties need to acquire the public keys first before establishing the pairwise keys. Therefore, the complexity comes from the operations in the public key request module and the key derivation module. On the other hand, the complexity in a V2I communication is dominated by the operations of the key derivation module.

To determine the number of CPU cycles, we insert the read timestamp counter and processor (rdtscp ID) instruction in the C programs of the public key request module and the key derivation module. Both programs are executed for 1000 times on a 32 bit Debian Linux operating system running on an Intel Core i7-2620M processor workstation. Table IX shows the average number of CPU cycles needed for both V2V and V2I communication. Simulation results show that the V2V communications require 70.84 megacycles to complete which is about 57.4% higher than the V2I communications. The increase is due to the additional operations to acquire the public keys in the public key request module. When our scheme is implemented on an OBU with a processor @500MHz [26], the computation time for the V2I and V2V communication scenarios can be calculated as follows:

$CT = clock\ cycles\ for\ a\ program/clock\ rate$ = 58.60 ms and 141.68 ms respectively, which fulfills the latency requirements of typical VN applications. These results demonstrate the feasibility of our scheme in the real world settings.

TABLE IX
NUMBER OF CYCLES FOR C IMPLEMENTATION OF SA-KMP

| Types of Comm. | Number of CPU Cycles in megacycles | | Total CPU cycles in megacycles | OBU @500MHz |
|---|---|---|---|---|
| | Public Key Request Module | Key Derivation Module | | Computational time (CT) |
| V2I / I2V | - | 29.30 | 29.30 | 58.60ms |
| V2V / I2I | 41.54 | 29.30 | 70.84 | 141.68ms |

VII. CONCLUSION

In this paper, we have proposed the SA-KMP scheme that leverages on the PKR scheme and the 3D matrix key agreement scheme to reduce the latency and the complexity of a certificate based PKI. The PKR scheme eliminates the complex certificate verification process while the 3D matrix key agreement scheme establishes symmetric keys to replace the expensive asymmetric cryptography. Through numerical analysis, we have shown that SA-KMP scheme is more efficient and scalable than the certificate based PKI scheme in terms of the network latency and transmission overhead, albeit a higher storage requirement. When compared to the PKR scheme, the transmission overhead and the storage overhead of SA-KMP is higher because of the extra authentication and the storing of RTA's signature to combat DoS attacks and collusion attacks respectively. Nevertheless, the communication latency of the SA-KMP is still lower than the PKR scheme. In addition, despite the higher storage cost, the storage requirement is still well below the storage limit of a modern OBU with 8GB storage space [26]. Besides, we have shown that the SA-KMP scheme is highly configurable in terms of establishing pairwise keys. We have also demonstrated that under the DoS attacks, the SA-KMP scheme outperforms the ECDSA based schemes with a lower authentication delay. Lastly, the SA-KMP scheme is verified to be robust against a broad range of attacks using both formal and informal verifications. For future works, we plan to extend our scheme to establish group keys among vehicles without the help of an RSU and look into the privacy related issues.

ACKNOWLEDGMENT

The authors would like to thank Prof. Rida Khatoun and the anonymous reviewers for their suggestions on this work.

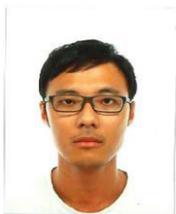
Hengchuan Tan received his BE degree in Electrical and Electronic Engineering from Nanyang Technological University in 2011. He is currently working towards his joint Ph.D. degree with Nanyang Technological University and Telecom ParisTech. His research interests include wireless network security and vehicular networks.

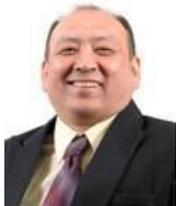
Dr. Maode Ma received his Ph.D. degree in Department of Computer Science from Hong Kong University of Science and Technology in 1999. Now, Dr. Ma is an Associate Professor in the School of Electrical and Electronic Engineering at Nanyang Technological University in Singapore. He has extensive research interests, including network security and wireless networking. Dr. Ma has over 300 international academic publications, including more than 130 journal papers and over 160 conference papers. Dr. Ma is the Fellow of *IET* and a senior member of *IEEE Communication Society and IEEE Education Society*. He is the Chair of the *IEEE Education Society*, Singapore Chapter. He is serving as an *IEEE Communication Society Distinguished Lecturer* from 2013 to 2016.

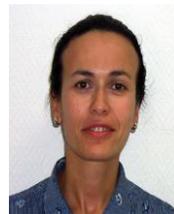
Houda Labiod is a Professor at Department INFRES (Computer Science and Network department) at Telecom ParisTech (previously named ENST) in Paris (France). In 2005, she obtained her HDR (Habilitation à diriger les recherches). Prior to this, she held a research position at Eurecom Institute, Sophia-antipolis, France. She obtained her Ph.D. thesis from the University of Versailles Saint-Quentin-en- Yvelines (France) in 1998. Her current research interests include wireless networks and autonomous networks (WLANs/ MANETs/ WSN/ NEMO/ Mesh/ vehicular), QoS, performance evaluation, security in intelligent transportation systems and link adaptation mechanisms. Dr. Labiod has published six books and many research papers in these areas. She is a Founder of IFIP NTMS Conference, on New Technologies, Mobility, and Security (NTMS2007). She served as an Associate Editor and a member of the Editorial Board for several journals.

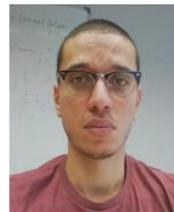
Aymen Boudguiga got a Telecommunications Engineering diploma from SUPCOM, Tunisia, in 2008. Then, he obtained a Ph.D. from Institut Mines-Telecom and the University of Pierre and Marie Curie, France, in 2012. In 2013, he worked as a researcher in the Communicating Systems Lab at the Atomic and Alternative Energy Commission (CEA), France. Then, he joined the Institute for Technological Research SystemX, France, in 2014. His research areas are key management, authentication, and risk analysis methods application to wireless networks and intelligent transportation systems.

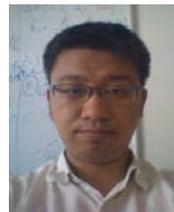
Jun Zhang received his BEng (2002) in computer science and technology from Shanghai Jiaotong University, and Ph.D. (2007) in computer science from Hong Kong University of Science and Technology. He was a research fellow at the Hong Kong City University (2007-2009), the Hong Kong Polytechnic University (2009-2011), a postdoctoral fellow at the Hong Kong University of Science and Technology (2011-2013), and currently a postdoctoral researcher in Telecom ParisTech. His research interests include wireless multi-hop networks (ad hoc/sensor/mesh networks), wireless LAN and vehicular networks.

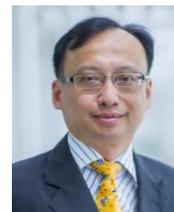
Dr. Peter Han Joo Chong is currently a Professor and Head of Department of Electrical and Electronic Engineering at Auckland University of Technology, Auckland, New Zealand. He received the Ph.D. degree in Electrical and Computer Engineering from the University of British Columbia, Vancouver, BC, Canada, in 2000. He is an Adjunct Faculty at the Department of Information Engineering, Chinese University of Hong Kong. He was previously an Associate Professor (tenured) from 2009 to 2016 and Assistant Professor from 2002 to 2009 in the School of Electrical and Electronic Engineering at Nanyang Technological University (NTU), Singapore. From 2001 to 2002, he was a Research Engineer at Nokia Research Center, Helsinki, Finland. Between 2000 and 2001, he worked in the Advanced Networks Division at Agilent Technologies Canada Inc., Vancouver, BC, Canada. He is the Co-Founder of P2 Wireless Technology based in Hong Kong. He is an Editorial Board Member of Security and Communication Networks and KSII Transactions on Internet and Information Systems. His research interests are in the areas of mobile communications systems including radio resource management, multiple access, MANETs/VANETs, multihop cellular networks and Internet of Things/Vehicles.